%% file: cerfpes_resub3.tex
\documentclass[aps,prl,amsmath,twocolumn,showpacs,superscriptaddress]{revtex4-1}

\usepackage{color}

\usepackage{graphicx}
\usepackage{hyperref}

\newcommand{\MHz}{\,\text{MHz}}
\newcommand{\GHz}{\,\text{GHz}}

\newcommand{\meV}{\,\text{meV}}
\newcommand{\eV}{\,\text{eV}}

\newcommand{\V}{\,\text{V}}
\newcommand{\kV}{\,\text{kV}}

\newcommand{\rad}{\,\text{rad}}
\newcommand{\mm}{\,\text{mm}}
\newcommand{\um}{\,\mu\text{m}}
\newcommand{\nm}{\,\text{nm}}

\newcommand{\fs}{\,\text{fs}}
\newcommand{\ps}{\,\text{ps}}

\newcommand{\W}{\,\text{W}}

\newcommand{\GWcm}{\,\text{GW/cm$^{2}$}}
\newcommand{\Vperps}{\eV/\text{ps}}

\newcommand{\degrees}{^\circ}
\newcommand{\urad}{\,\mu \text{rad}}
\newcommand{\mrad}{\,\text{mrad}}

\newcommand{\rfcavitymode}{TM$_{020}$}

\newcommand{\Woneoneone}{W(111)}
\newcommand{\Ekin}{E_\text{$\Delta\phi$}}
\newcommand{\EkinI}{E_\text{$\Delta$I}}
\newcommand{\figref}[1]{Fig.~\ref{#1}}

\newcommand{\stanfordphysics}{\affiliation{Physics Department, Stanford University, Stanford, California 94305, USA}}

\newcommand{\usfphysics}{\affiliation{Department of Physics and Astronomy, University of San Francisco, 2130 Fulton St., San Francisco, CA 94117, USA}}

\begin{document}
\title{Ultrafast time resolved photo-electric emission}
\author{Thomas Juffmann}\email{Corresponding author: juffmann@stanford.edu}\stanfordphysics
\author{Brannon B. Klopfer}\stanfordphysics
\author{Gunnar E. Skulason}\stanfordphysics
\author{Catherine Kealhofer}\stanfordphysics
\author{Fan Xiao}\stanfordphysics
\author{Seth M. Foreman}\usfphysics
\author{Mark A. Kasevich}\stanfordphysics

\begin{abstract}
The emission times of laser-triggered electrons from a sharp tungsten tip are directly characterized under ultrafast,  near-infrared laser excitation at Keldysh parameters $6.6< \gamma < 19.1$.  Emission delays up to $10\fs$ are observed, which are inferred from the energy gain of photoelectrons emitted into a synchronously driven microwave cavity. $\sim \fs$ timing resolution is achieved in a configuration capable of measuring timing shifts up to $55 \ps$. The technique
can also be used to measure the microwave phase inside the cavity with a precision below $70\fs$ upon the energy resolved detection of a single electron.
\end{abstract}

\maketitle
Ultrafast laser excitation of metallic nanostructures has recently been exploited to generate triggered ultrafast electron sources \cite{hommelhoff_ultrafast_2006, hommelhoff_field_2006,barwick_laser-induced_2007, ropers_localized_2007}.   These sources are finding increasing application in ultrafast electron microscopy and electron diffraction \cite{hoffrogge_tip-based_2014, gulde_ultrafast_2014} as well as in the characterization of ultrafast electronic \cite{kealhofer_ultrafast_2015} or optical \cite{wimmer_terahertz_2014} signals. They are further discussed as possible electron sources for laser-based electron accelerators \cite{breuer_laser-based_2013, peralta_demonstration_2013}.   While the basic physical processes underlying ultrafast photo-excitation mechanisms have been understood for decades \cite{yen_1980}, a detailed experimental understanding of these mechanisms and their interplay in technologically relevant systems is still lacking. 

Recent experiments show that prompt electron emission processes contribute to the electron signal \cite{hommelhoff_ultrafast_2006, kruger_attosecond_2012, bormann_tip-enhanced_2010} and 2-photon pump-probe experiments provide evidence for the emission being restricted to less than $100\fs$ \cite{barwick_laser-induced_2007}. However electron energy relaxation within the metal tip and barrier tunneling dynamics offer paths for delayed emission on timescales ranging from a few $\fs$ up to $1 \ps$ \cite{yanagisawa_energy_2011, kealhofer_ultrafast_2012}. In prior work, such delayed emission has been inferred indirectly from measurements of the energy or momentum distribution of emitted electrons \cite{yanagisawa_energy_2011, Sekatskii_2001}. A detailed physical model is surprisingly complex, since the anticipated timescales for electron relaxation due to electron-electron collisions is comparable with the excitation timescales associated with the energy transfer to the electrons and the timescales for below barrier tunneling \cite{pant_2013}.  

In this Letter we directly measure relative emission delays for photo-excited electrons from sharp metal tips in a technologically relevant regime, where near-infrared ($\sim 775$ nm) laser excitation with $\sim10\fs$ pulse duration at Keldysh parameters $6.6<\gamma<19.1$ leads to a current of up to $0.1$ electrons per laser pulse ($\gamma \equiv \omega\sqrt{2m\varphi_\text{eff}/(qF)}$, where $m$ and $q$ are the mass and the charge of the electron, $\omega$ is the mean angular laser frequency, $F$ is the laser induced, locally enhanced, electric field at the tip apex and $\varphi_\text{eff}$ is the effective work function of the material \cite{Keldysh_1965}). Electron emission delays are characterized by measuring the energy shift of electrons which photo-emit into a strong microwave field. As we show below, synchronization of the microwave field phase with respect to the ultrafast laser pulse used to photo-excite electrons allows for $\fs$ resolution of the emission time with a timing dynamic range limited to half the period of the microwave field ($\tau_{RF}/2\sim55\ps$). This technique avoids interference effects between a pump and probe pulse \cite{petek_femtosecond_1997} as well as probe induced quiver motion and electron recollisions \cite{krausz_2009}. Locating the tip inside the microwave cavity leads to field enhancement of the microwave field and avoids dispersion of the electron beam prior to interaction with the microwave field.

\begin{figure}
\includegraphics[width=\columnwidth]{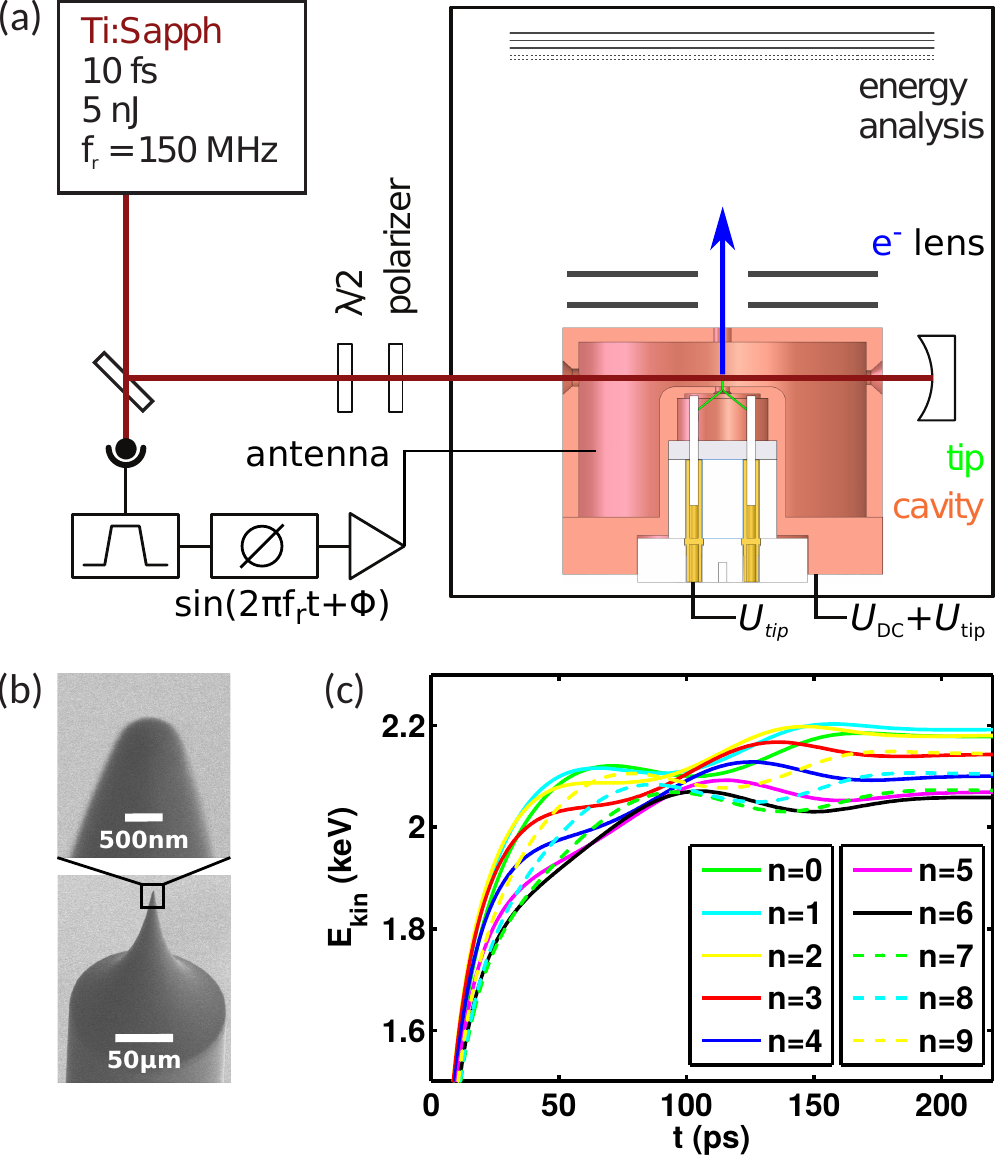}
\caption{(color online). (a) Schematic of the apparatus.  Shown is a cross-section of the cylindrically symmetric re-entrant microwave cavity. The distance from the tip to the exit plane is $4.25\mm$. The tip is biased to a voltage of  $U_\text{tip}=-35\V$, with respect to ground, and the cavity is floated to a voltage of $U_\text{DC}+U_\text{tip}$. (b) Scanning electron microscope pictures of the tungsten nanotip show a radius of $r=400 \nm$ and a cone angle of $\sim17\degrees$. (c) Simulated energy modulation at a bias voltage of $2.1\kV$ as a function of time and initial microwave phase $\phi_n=2\pi n/10\rad$. At this bias voltage the electrons leave the cavity after approximately 1.5 periods of the microwave field, which is why the energy modulation curves become flat after  $\sim 165\ps$.\label{fig:schematicetc}}
\end{figure}

The measurement approach is illustrated in \figref{fig:schematicetc}a. A titanium-sapphire laser provides $10\fs$ short laser pulses at a repetition rate of $\sim 150\MHz$. The incident light is focused to a waist of $6\um$ ($1/e^2$ radius) onto a sharp tungsten tip  [\Woneoneone, tip radius $r\sim400\nm$, see \figref{fig:schematicetc}b], which is voltage biased with a DC voltage $U_{DC} \sim 2$ kV. The laser polarization is oriented parallel to the tip axis.  The tungsten tip is located inside a cylindrically symmetric re-entrant cavity. Along the symmetry axis the transverse components of the electric field of the \rfcavitymode{} mode are zero and the longitudinal electric field is homogeneous except for the local field enhancement close to the tip. The cavity has an entry and an exit for the laser beam as well as an exit aperture for the electron pulses. The cavity has a resonance frequency $f_{RF} =  9.08\GHz$ and \emph{Q} of $\sim 2500$. 

After laser-induced emission, electrons are accelerated by the DC bias field and the microwave field, both of which are enhanced close to the tip. The interaction with the microwave field leads to a dependence of the final kinetic energy of the electrons on the phase of the microwave field at the moment the electrons leave the tip. After exiting the cavity the electrons are collimated using two lens electrodes and then decelerated towards a retarding field energy analyzer with a resolution of $\sim 1.5\eV$ (full width, half maximum).  In order to synchronize the microwave field phase with the incident laser pulse,  we derive the cavity microwave signal directly from a harmonic of the laser repetition rate, which is obtained from a fast photodiode which detects a reference beam of the laser (see \figref{fig:schematicetc}a). The photodiode signal is filtered, phase shifted, and amplified before being coupled into the cavity (see supplemental material for details).

\figref{fig:schematicetc}c shows charged particle tracing simulations of electron energy as a function of time at a bias voltage $U_\text{DC} = 2.1\kV$. The electrons gain $\sim$ 95\% of their kinetic energy within the first $50\ps$ following emission. The sensitivity of this technique is maximal if the electrons exit the cavity after interaction times close to half-integer multiples of the microwave period, which can be controlled by tuning  $U_\text{DC}$.  

\begin{figure}
 \includegraphics[width=\columnwidth]{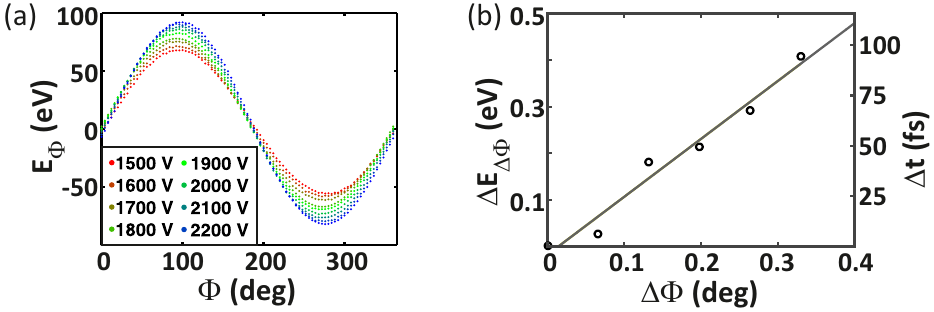}
\caption{(color online). (a) Measured energy modulation $E(\phi)$ as a function of initial microwave phase $\phi$ for different bias voltages $U_\text{DC}$. (b) Varying the microwave phase (see supplemental material) at the zero crossing of the energy modulation curve in (a) shows the accuracy of the technique: Averaging 15 measurements of $\Delta \Ekin$ (see text)  yields a standard error of the residuals from a linear fit of $2.1\fs$. \label{fig:data}}

\end{figure}

\figref{fig:data}a shows the measured electron energy gain $E(\phi)$ as a function of the microwave phase $\phi$. There is a unique phase to energy mapping for phase shifts smaller than $\pi$. This corresponds to a dynamic range of emission time measurements of $\Delta t_\text{max}=\tau_\text{RF} /2 = 55 \ps$ with a sensitivity given by the slope of the energy modulation curve.  At $0.8\W$ of microwave input power the final electron energy shows a peak-to-peak modulation of $\sim 164\eV$ at $U_\text{DC} = 2.1\kV$. The slope is $\sim 4.6\Vperps$ at the zero crossing of this modulation curve.  All following measurements were performed in a $10 \eV$ energy window centered at the zero crossing of the energy modulation curve, corresponding to a dynamic range of about $2.5 \ps$. No significant electron flux was observed outside of this energy window.

\figref{fig:data}b shows the measured timing resolution of the apparatus.  Each data point represents the average of 15 measurements of the quantity 
\begin{eqnarray*}
\Delta \Ekin &\equiv&\frac{1}{2}
\left[
	(E(\phi_0 + \Delta \phi)
	-
	E(\phi_0 + \Delta \phi + \pi))
	\right.\\
	&&
	-\left.
	(E(\phi_0)
	-
	E(\phi_0 + \pi))
\right]
,
\end{eqnarray*}
where $\Delta \phi$ is controlled using a direct digital synthesizer and $\phi_0$ is chosen to be near a zero crossing of the energy modulation curve shown in \figref{fig:data}a.  Measurements at both the positive and negative slopes suppress possible errors from drifts in $E(\phi_0)$, due to, for example, slow changes in the tip work function. The energy depends on phase linearly and the deviations from the linear fit show a standard error corresponding to $2.1\fs$ (see supplemental material for additional accuracy analysis).

In order to study electron emission delays, we compare electron emission energy at two differing laser intensities.  Specifically, we determine the quantity 

\begin{eqnarray*}
\Delta \EkinI &\equiv&\frac{1}{2}
\left[
	(E(\phi_0,\text{I})
	-
	E(\phi_0 +\pi,\text{I}))
	\right.\notag\\&&\left.
	-
	(E(\phi_0,\text{I$_0$})
	-
	E(\phi_0 + \pi,\text{I$_0$}))
\right]
,
\end{eqnarray*}
where $I_0 = 295\GWcm$ (our highest intensity). Note that $I$ denotes the intensity of the incoming laser pulse and that the optical fields are enhanced close to the tip \cite{martin_2001, kruger_attosecond_2012}. For the tungsten tip used in this study an optical field enhancement factor of 1.5 is expected \cite{Hommelhoff_2015} from numerically solving Maxwell's equations using finite-difference time-domain  (FDTD)  calculations \cite{Thomas_2015}. 

\begin{figure}
\includegraphics[width=\columnwidth]{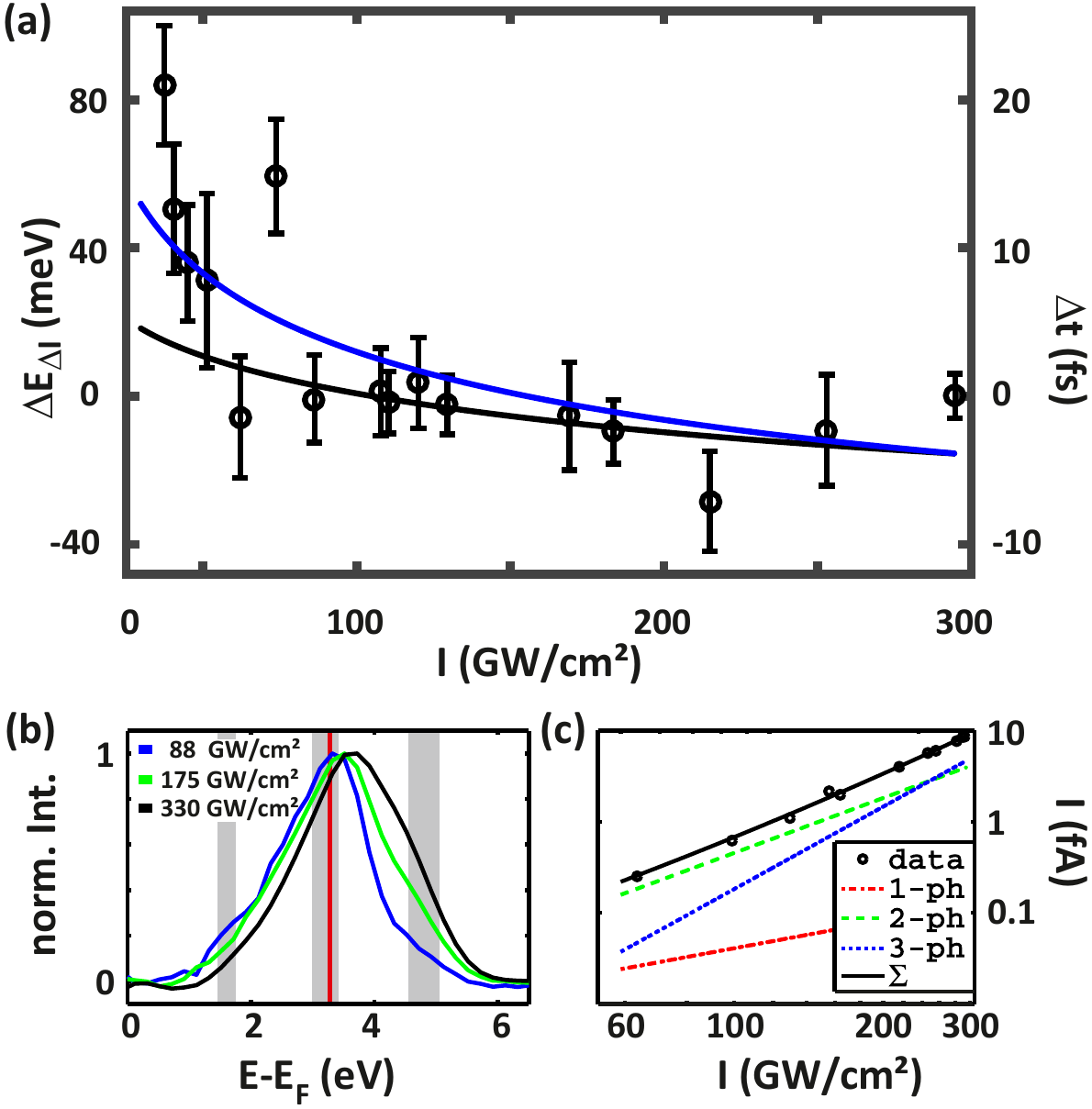}
\caption{(color online). (a) $\Delta \EkinI$ in a regime of Keldysh parameters of $19.1>\gamma>6.6$. The data is fitted with a model based on the Keldysh time (blue line, see text). The kinematic contribution to the observed energy shift is indicated by the black line. A delay is only observed for low laser intensities $< 75\GWcm$. (b) Normalized electron energy spectra for various laser intensities. The spectra are referenced to the Fermi energy $E_{F}$. The red line shows the effective work function $\varphi_\text{eff}=\varphi_\text{0}-\sqrt{\frac{q^{3}U_{DC}}{4\pi\epsilon_{0}kr}}\sim3.27\eV$, where $\varphi_{0}=4.5\eV$ is the work function of \Woneoneone \cite{todd_1973} and $k\sim5$ is a geometry dependent parameter, which influences the local field enhancement of the DC field \cite{gomer_1961}. The gray shaded areas are the energies provided by $n$ absorbed photons ($n=1,2,3$). Their width is given by the spectral width of the titanium-sapphire laser ($\sim0.3\eV$). Note that the instrument response of the energy analyzer is $1.5\eV$ wide (FWHM). We find an overall shift to higher energies at higher laser intensities. (c) A polynomial fit (black line) to the totally emitted current as a function of laser intensity gives the relative contribution of $n$-photon induced emission. While at low laser intensity the signal is dominated by 2-photon induced emission, 3-photon induced emission becomes more significant at higher laser intensities. 
\label{fig:data2}}
\end{figure}

We measure $\Delta \EkinI$ for laser intensities between $35\GWcm$ and $295\GWcm$, corresponding to a Keldysh parameter of $19.1>\gamma>6.6$. The measured energy shifts are shown in \figref{fig:data2}a, where every data point represents the average of at least 50 individual delay measurements. 

Large energy shifts are observed at low laser intensities.  These shifts are due to delayed photoemission as well as due to the kinematics of the free electrons.  The kinematic shifts are due to energy dependent interaction times between the electrons and the microwave field.  According to charged particle tracing simulations, it amounts to $52\meV$ for electrons whose kinetic energy initially differ by one photon energy ($h\nu=1.6 \eV$).  Note that this shift is much smaller than the initial kinetic energy difference, due to the differential measurement scheme involving measurements at $\phi_0$ and at $\phi_0 + \pi$. Independent determination of the kinematic contribution, in combination with the measured total delay, thus allows for determination of the photoemission delay.


We quantify the kinematic contribution by characterizing the electron emission energy as a function of laser intensity. \figref{fig:data2}b shows electron energy spectra taken at several average laser intensities with no microwave field in the cavity. The recorded spectra are referenced to the Fermi level and suggest multiphoton induced electron emission with a main contribution at an energy of  $2h\nu$ and a growing above barrier contribution at energies of $3h\nu$ at higher laser intensities. This can also be seen in \figref{fig:data2}c, which shows the detected electron current as a function of laser intensity. A polynomial fit (black line) to the data reveals the respective contributions of $n$-photon induced emission in agreement with earlier studies that showed quantized energy spectra \cite{yanagisawa_energy_2011}  and intensity law dependencies of the emitted current \cite{barwick_laser-induced_2007, ropers_localized_2007, hilbert_2009} at comparable laser intensities. With the polynomial fit in \figref{fig:data2}c and the results from charged particle tracing simulations the laser intensity dependent kinematic energy shift can be determined, as shown in \figref{fig:data2}a.

The additional energy shift observed in the experiment is proportional to a delay in emission. For the following discussion the measured energy shift is converted into a timing delay, taking into account the measurement sensitivity of $\sim4\Vperps$. For $I>75\GWcm$ no statistically significant delay is observed. Assuming that emission at the highest (reference) laser intensity is largely prompt we infer that, for $I>75\GWcm$, laser triggered photoemission from tungsten nanotips provides excellent timing resolution.

For lower laser intensities, we observe substantial delays of up to $10\fs$. Recent theory \cite{Orlando_2014} has proposed that tunneling delays are proportional to the Keldysh time $\tau_K=\gamma \tau_L/2\pi$ \cite{Keldysh_1965}, where $\tau_L$ is the laser period. Fitting the simple model $\Delta t = O+C(\tau_K(I)-\tau_K(I_0))+\Delta t_{kin}$, where $\Delta t_{kin}$ is the kinematic contribution, yields the blue curve in \figref{fig:data2}a (see supplemental material for fitting details). The fitted proportionality constant of $C= 1.43$ is in good agreement with a recently reported value \cite{Yanagisawa_2014}. The fitted offset of $3.9\fs$ is likely due to the statistical repeatability of each measurement. Other theoretical work predicts timing shifts of a few fs in a similar system, where the effective  binding energy of an electron is close to a multiple of the photon energy \cite{pant_2013}. Further studies at higher energy resolution will be required to clarify whether the deviations from the simple Keldysh model may be attributed to such resonances. Such a model, based on solving the time-dependent Schr\"odinger equation, will also be more appropriate when the photoemission delays approach the duration of the laser pulse.



The sensitivity of the described apparatus could also be used to measure the microwave phase within the cavity. At $8\W$ of microwave input power the sensitivity of the technique was $7.8\Vperps$. Assuming perfect microwave to laser synchronization, and given the intrinsic width of the electrons' energy distribution of about $0.5\eV$ one could measure the phase of the microwave field to within $64\fs$ (modulo $\tau_\text{RF}/2$), or $0.6\mrad$ (modulo $\pi$) at $9.08\GHz$, upon the energy resolved detection of a single electron. Such precision in measuring the microwave phase in situ, i.e. within the microwave cavity itself \cite{Brussaard_Direct_2013}, will directly benefit techniques such as temporal focusing \cite{van_oudheusden_compression_2010, gliserin_compression_2012, fill_sub-fs_2006, chatelain_ultrafast_2012, kassier_photo_2012} and aberration-free lensing \cite{pasmans_microwave_2013} which compress or maintain femtosecond electron pulse durations at samples distant from the source.

We have directly characterized the timing of laser triggered electron emission by measuring the energy gain/loss of photoemitted electrons in a microwave cavity. While microwave fields represent a minimally invasive probe avoiding quiver motion of the electron, recollision processes or interference of the pump and probe pulse, they still allow for  $\fs$ resolution and a dynamic range extending to $55 \ps$. We find that electron emission is prompt for Keldysh parameters less than $\sim 13$, indicating that laser triggered electron sources have excellent timing resolution in this regime.  At larger Keldysh parameters, we observe significant (as large as $10\fs$) delays.

\begin{acknowledgements}
TJ would like to thank Dominik Leuenberger and Philipp Haslinger for helpful discussions. This research is funded by the Gordon and Betty Moore Foundation, and by work supported under the Stanford Graduate Fellowship.
\end{acknowledgements}

\bibliographystyle{apsrev4-1}
\bibliography{cerfpes}

\newpage
\newpage
\input{supplementary_material.tex}
\end{document}

%% file: supplementary_material.tex
\newpage
\noindent \textbf{Supplemental Material\newline}

\begin{figure}[h!]
 \includegraphics[width=\columnwidth]{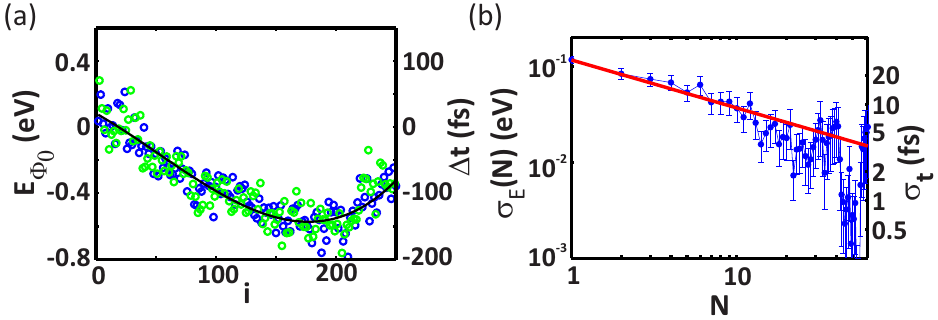}
\caption{(color online). (a) Repeated measurements of $E(\phi=\phi_0)$ at the zero crossing of the energy modulation curve at $U_\text{DC} = 2.1\kV$. The maximum slope of the third order polynomial fitted to the data (solid line) indicates a microwave phase drift corresponding to $< 4\fs/\text{min}$. The drift corrected measurements have a standard deviation corresponding to $22\fs$. Averaging yields a resolution of cavity enhanced microwave photoelectron streaking on the single $\fs$ level. This is confirmed in (b), where the Allan deviation of a differential measurement $\Delta E=E_\text{blue}(\phi_0)-E_\text{green}(\phi_0)$ is plotted as a function of the number of measurements $N$. Good agreement is observed with the statistical error $\propto1/\sqrt{N}$ (solid line).  \label{fig:S2}}
\end{figure}

\noindent \textbf{Raw data and additional measurement resolution analysis:} Repeated measurements of $E(\phi=\phi_0)$ at the zero crossing of the energy modulation curve at $U_\text{DC} = 2.1\kV$ are shown in \figref{fig:S2}a. Correcting for a slow drift  $< 4\fs/\text{min}$ yields a standard deviation of $22\fs$ for a single measurement. To test for the resolution of cavity enhanced microwave photoelectron streaking, the dataset is divided into two interleaved datasets (alternating blue and green data points) and Allan deviations for the differential quantity $\Delta E=E_\text{blue}(\phi_0)-E_\text{green}(\phi_0)$ are calculated and shown in \figref{fig:S2}b. It agrees well with the standard deviation of the sample mean and shows that a resolution of a few $\fs$ can be reached with sufficient averaging.

\noindent \textbf{Tunneling delay data analysis:} At the reference intensity $I=I_0$ an additional data point is introduced at $\Delta t = 0\fs$ with an error bar of $\sigma_{\Delta t}(I=I_0)=\sigma_{\Delta t}(I=I_1)/\sqrt{2}=1.5 \fs$, where $I_1$ is the intensity at the adjacent data point. This restricts the fitting of the offset to within the accuracy of the measurement. Without this data point the proportionality constant and offset would be $C=2.75$ and $O=7.4\fs$, respectively. All data points are weighted by the variance of each measurement. 

\begin{figure}
 \includegraphics[width=\columnwidth]{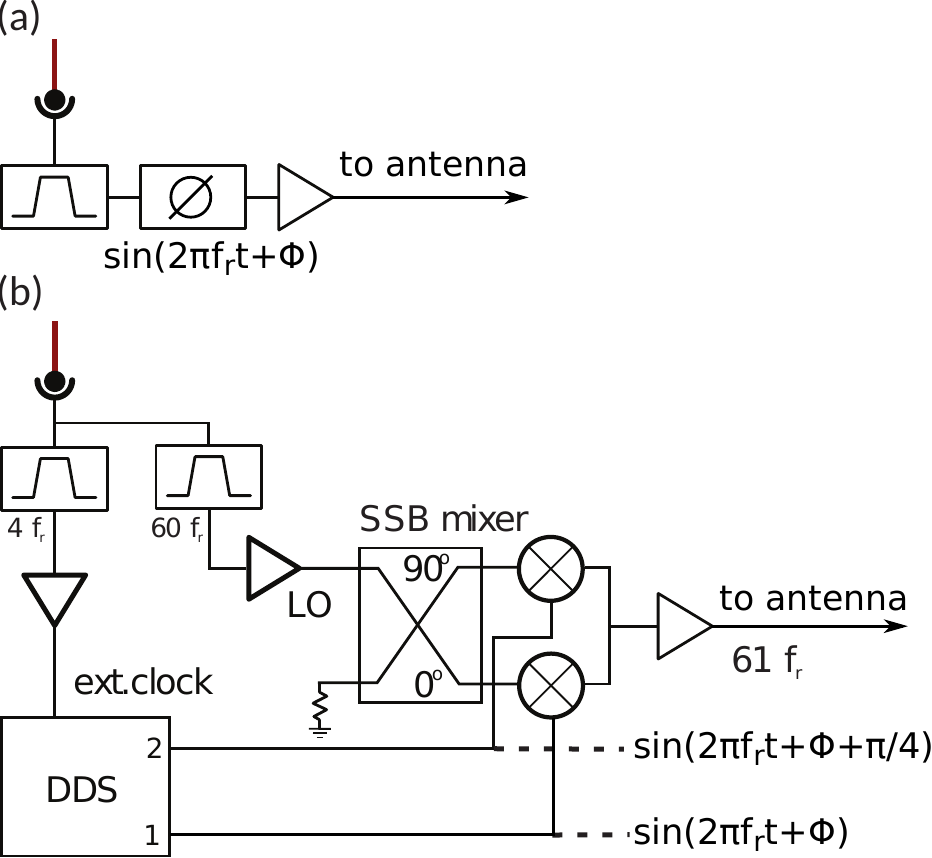}
\caption{(To synchronize the microwave field with the laser pulses a pick off beam is detected with a fast photodiode. (a) For most measurements, the signal at the 61\textsuperscript{st} harmonic of the laser repetition rate is filtered, phase shifted, amplified and fed into the cavity. (b) For measurements that require fast switching between the positive and negative zero-crossing of the energy modulation curve, the phase shifting was achieved using a direct digital synthesizer and a single-sideband mixer (see text). \label{fig:S1}}
\end{figure}
\vspace{6 pt}
\noindent \textbf{Synchronization of microwave fields:} In order to synchronize the microwave fields inside the cavity with the laser pulses, we derive the microwave signal directly from a fast photodiode (Hamamatsu G4176), which detects a pick-off beam of the laser. There are two ways we manipulate the microwave field before we couple it into the cavity. For the data shown in \figref{fig:S2}a and \figref{fig:S2}b the signal at the 61st harmonic of the laser repetition rate is filtered, phase shifted using an electromechanical phase shifter, amplified and coupled into the microwave cavity using an antenna (see \figref{fig:S1}a). All other data was taken using the microwave circuit shown in \figref{fig:S1}b, which allows alternating quickly between measurements at the positive and the negative zero crossing of the energy modulation curve shown in \figref{fig:data}a by replacing the manually operated phase shifter with a direct digital synthesizer (DDS, Novatech 409B with 14-bit resolution corresponding to $0.022\degrees$ ($384\urad$) or a timing shift of $6.722\fs$ at $9.08\GHz$). The photodiode signal is divided into a low frequency component at the 4\textsuperscript{th} harmonic of the laser repetition rate and a high frequency component at the 60\textsuperscript{th} harmonic of the laser repetition rate. The low frequency component is used as a clock for the DDS, which generates a signal at the repetition rate of the laser with a digitally controllable phase. This phase shifted signal is single-sideband (SSB) mixed with the high frequency component to create a signal at the 61\textsuperscript{st} harmonic, which is at the cavity resonance. It is amplified, filtered and coupled into the cavity.